\documentclass[aps,epsfig,article,superscriptaddress]{revtex4}
\usepackage{dcolumn}
\usepackage{graphicx}
\usepackage{amsmath}
\usepackage{latexsym}
\usepackage{amsfonts}
\usepackage{amssymb}
\usepackage{array}
\usepackage{color}
\usepackage{subfigure}
\usepackage{verbatim}
\usepackage{blindtext}
\usepackage{mathtools}
\usepackage[normalem]{ulem}
\usepackage{appendix}
\usepackage{chngcntr}
\newcommand{\ket}[1]{\left\vert#1\right\rangle}
\newcommand{\bra}[1]{\left\langle#1\right\vert}

\begin{document}
\newcommand{\Q}[1]{{\color{red}#1}}
\newcommand{\blue}[1]{{\color{blue}#1}}
\newcommand{\red}[1]{{\color{red}#1}}
\newcommand{\green}[1]{{\color{green}#1}}
\newcommand{\Eq}[1]{\begin{align}&#1\end{align}}
\title{
Slowing quantum decoherence of oscillators by hybrid processing }

\author{
Kimin Park}
\affiliation{Center for Macroscopic Quantum States (bigQ), Department of Physics, Technical University of Denmark, Building 307, Fysikvej, 2800 Kgs. Lyngby, Denmark}
\affiliation{Department of Optics, Palacky Univeristy, 77146 Olomouc, Czech Republic}
\author{Jacob Hastrup}
\affiliation{Center for Macroscopic Quantum States (bigQ), Department of Physics, Technical University of Denmark, Building 307, Fysikvej, 2800 Kgs. Lyngby, Denmark}
\author{Jonas Schou Neergaard-Nielsen}
\affiliation{Center for Macroscopic Quantum States (bigQ), Department of Physics, Technical University of Denmark, Building 307, Fysikvej, 2800 Kgs. Lyngby, Denmark}
\author{Jonatan Bohr Brask}
\affiliation{Center for Macroscopic Quantum States (bigQ), Department of Physics, Technical University of Denmark, Building 307, Fysikvej, 2800 Kgs. Lyngby, Denmark}
\author{Radim Filip}
\affiliation{Department of Optics, Palacky Univeristy, 77146 Olomouc, Czech Republic}
\author{Ulrik L. Andersen}
\affiliation{Center for Macroscopic Quantum States (bigQ), Department of Physics, Technical University of Denmark, Building 307, Fysikvej, 2800 Kgs. Lyngby, Denmark}
\date{\today}

\begin{abstract}
Quantum information encoded into superposition of coherent states is an illustrative representative of practical applications of macroscopic quantum coherence possessing.
However, these states are very sensitive to energy loss, losing their non-classical aspects of coherence very rapidly. 
An available deterministic strategy to slow down this decoherence process is to apply a Gaussian squeezing transformation prior to the loss as a protective step. 
Here, we propose a deterministic hybrid protection scheme utilizing strong but feasible interactions with two-level ancillas immune to  spontaneous emission. 
We verify robustness of the scheme against  dephasing of qubit ancilla. 
Our scheme is applicable to complex superpositions of coherent states in many oscillators, and remarkably, the  robustness to loss is \emph{enhanced} with the amplitude of the coherent states.
This scheme can be realized in experiments with atoms, solid-state systems and superconducting circuits. 
\end{abstract}

\maketitle

\section{Introduction}

Quantum information processing based on continuous-variable (CV) resources ~\cite{BraunsteinRMP2005CV,Cerf2007CVBook,WeedbrookRMP2012CV,AndersenNatPhys2015Hybrid,Lvovsky2020nongaussian} represents an interesting alternative to a more common discrete-variable (DV) approach based on photons ~\cite{KokRMP2007PhotQub, Obrien2009Phot,Flamini2018}, or other particles~\cite{Bruzewicz2019APRtrappedion,Kjaergaard2020SuperCondQub,SemiconductorSpins2019}.
In particular, non-Gaussian states represented by the superposition or entanglement of a finite number of coherent states are useful resources in various protocols in both approaches~\cite{SandersRev2012ECS}. 
The interest in these states has been fueled by  various schemes for  sensing~\cite{Zurek2001Compass, JooPRL2011ECS},  computing~\cite{RalphPRA2003, JeongPRA2002, MarekPRA2010Catgates,MirrahimiNJP2014cat} and communication  \cite{vanLoockPRL2006HybridRepeaterCoherent,SangouardRMP2011Repeater}  with promising merits in scalability and fault-tolerance \cite{LundPRL2008,LeghtasPRL2013,RosenblumScience2018parity}.
There have been significant efforts into the generation of the superposition of coherent states (SCSs) and entangled coherent states (ECSs), both in the optical \cite{GerritsPRA2010,Neergaard-NielsenPRL2010,OurjoumtsevNat2007,DongJosaB2014CatGeneration,HackerNatPho2019},  microwave  \cite{WinelandRMP2013,KienzlerPRL2016,DelegliseNature2008,HarocheRMP2013,VlastakisScience2013,PfaffNatPhys2017,WECSgeneration}  and phononic domains \cite{HoffPRL2016,KhoslaPRX2018,TehPRA2018}. 
Moreover, simple CV quantum processing tasks have been carried out on optical fields \cite{TipsmarkPRA2011HadamardCat,Larsen2020multimode}, while more advanced processing has been demonstrated in the microwave regimes including quantum error correction \cite{HeeresNatComm2017,OfekNature2016CatExperiment}. 


A formidable challenge associated with the faithful processing, transmission and storage of SCSs lies in the inevitable decoherence by bosonic loss.
While non-Gaussian \cite{AokiNatPhys2009,Lassen2010LossCorr} or correlated \cite{LassenPRL2013GaussianComm} noise sources can be circumvented for any type of encoding in bosonic modes by means of simple Gaussian transformations,  the stationary bosonic loss – the dominant decoherence source in most bosonic systems – is non-trivial to correct for. 
To overcome the effect of loss in quantum information processing, one could use the protocol of entanglement distillation of resources \cite{RalphPRA2011ECDistill} in combination with deterministic teleportation, or more advanced quantum error correction coding schemes  in which errors are corrected by encoding quantum information into special bosonic codes such as Gottesman-Kitaev-Preskill (GKP) states~\cite{OfekNature2016CatExperiment,Noh2019GKP, Fabre2019Combs,HeeresNatComm2017,Hastrup2019GKP,MichaelPRA2016Binomial, NohPRA2020SurfaceGKP, FluhmannHomeNature2019, SchoelkopfNature2020,TzitrinPRA2020GKP,TerhalQST2020GKP}, binomial states~\cite{MichaelPRA2016Binomial,HuNatPhys2019Binomial,noh2021quantum} or SCSs~\cite{PuriSciAdv2020Cat,GrimmNature2020KerrCat,chamberland2020building}. 
However, those ``ultimate'' approaches requires a challenging large resources to universal loss correction. 

Bosonic losses can be also partially but {\em deterministically} compensated for by transforming the encoded quantum state with a special symmetry into a state that is more robust against losses. 
For example, by conditionally de-amplifying a coherent state \cite{RalphPRA2011ECDistill,MullerPRA2012CloningCoherentstate,HawNatComm2016nocloningamplifier,BrewsterPRA2018SqzAmp}, or unconditionally squeezing their superposition states \cite{LeJeannicPRL2018,FilipJOB2001Amplifier,Serafini2004MinimumDecoherenceCat,
FilipPRA2013GaussianAdaptation} prior to a lossy bosonic channel, the coherence or non-classicality can survive for longer time. 
In these two protective schemes enabled by pre-processing, the quantum information has resided entirely in a bosonic quantum state.  

In this article, we propose a next step in  deterministic hybrid protection strategy for SCSs and ECSs, or a qubit \emph{bypass} strategy, where the quantum information of these states does not reside fully in the lossy bosonic mode - as in previous schemes - but is substantially converted into a lossless two-level system. 
The quantum information is by large protected from bosonic loss by  such a hybrid strategy,  albeit being traded for phase decoherence of the two-level system. 
Our protection scheme is enabled by a strong coherent coupling – the Rabi coupling – which can be implemented in superconducting systems, ion systems, or photonic or molecular crystal systems~\cite{NoriRabiRMP2019,SolanoRabiRMP2019,MuellerNature2020Plasmon,Fluhmann2018Rabi,
LangfordNatComm2017Rabi,
Lv2018RabiTrappedIon}  where the bosonic mode is represented by a microwave,  a phononic, or even an optical field.
In this scheme, the number of used two-level systems and Rabi couplings is kept at a minimum, while the coupling strength is dictated by the amplitudes of the coherent states of the SCSs and ECSs. 
Our approach is therefore complementary to an approach where the available number of Rabi gates and ancillas are unlimited which in principle would allow for a complete transfer of the information from the bosonic mode to the two-level systems~\cite{JacobDigitalizer}.
We analyze our protocol using different measures (i.e. fidelity, coherence in phase space and entanglement), with respect to the simple protection scheme of pre-squeezing \cite{LeJeannicPRL2018}.  
We find that  the robustness to losses is significantly increased in all these measures, remarkably  for large amplitude SCSs. 
Our strategy can be extended to states with higher complexity, e.g. with more coherent state components or modes, thereby indicating that further extensions of the method to general CV states might be viable. 

We proceed by introducing the hybrid protection protocol in section \ref{Sec:protool}, and presenting the results in section \ref{sec:results}. In section \ref{sec:conclusion} we conclude the study. 



\section{QUBIT Bypass protocol} 
\label{Sec:protool}





Our qubit bypass protocol is based on the quantum Rabi model~\cite{NoriRabiRMP2019,SolanoRabiRMP2019}.
This model incorporates a CV-DV hybrid non-Gaussian interaction, where an electro-magnetic field or mechanical oscillator strongly couples to a two-level system, e.g. an atom. 
The unitary evolution opeator associated with the coupling in the model is written as  $\exp[i t \hat{\sigma}_\theta \hat{X}_\phi]$ with strength $t$ (the strength of the Hamiltonian is combined with the duration time), where a Pauli operator $\hat{\sigma}_\theta$ with $\theta=\{x,y,z\}$ acts on the two-level system, and the generalized quadrature operator $\hat{X}_{\Phi}=\frac{\hat{a}e^{-i\Phi}+\hat{a}^\dagger e^{i\Phi}}{\sqrt{2}}$ with arbitrary angle $\Phi$ governs the field. 
The rotating-wave approximation is not valid within this model, and the unitary operator contains both rotating and counter-rotating terms.
Strong Rabi coupling has been achieved experimentally in cavity QED systems such as trapped ions~\cite{Lv2018RabiTrappedIon} or superconducting systems~\cite{Markovic2018RabiSC} among many other systems, and has been simulated in all-optical setup~\cite{ParkNJP2020Rabi}.
It can be used to create diverse classes of non-Gaussian effects such as nonlinear phase gates~\cite{ParkPRA2016, ParkNJP2018}, which are essential in CV quantum information processing~\cite{WeedbrookRMP2012CV}.






The unitary Rabi gate executes a controlled displacement operation, where the conjugate generalized momentum, $\hat{P}_\Phi=\hat{X}_{\Phi+\pi/2}$, of an oscillator is displaced depending on the state of an ancillary qubit encoded in $\hat{\sigma}_\theta$-eigenstates. 
Here, we will drop the subscript for $\Phi=0$ below for these operators.
The Rabi gate is particularly suitable for the generation of SCSs \cite{SolanoRabiRMP2019,NoriRabiRMP2019}. 
These controlled displacement operations have been used in numerous experiments to generate superpositions of two and four coherent states (for example, as in \cite{RoszakSciRep2015Decoherence}). 
In the current work, we will reverse the generation process by mapping the SCSs  onto a two-level ancilla, or more precisely,  partially converting the complex coefficients of SCSs onto the complex coefficients of a qubit, addressing the possibility to manipulate the SCSs by Rabi gates  for the first time. 

{\em Two coherent-state superposition---} 
Throughout the paper, we will refer the superposition of $n$ coherent states as n-SCSs.
We start with the simplest SCSs, namely the arbitrary superposition of two coherent states (2-SCS) of opposite amplitudes, $\ket{\psi_0}=N(\mu \ket{\alpha}+\nu \ket{-\alpha})$. 
The normalization factor is given by $N^{-2}=|\mu|^2+|\nu|^2+(\mu^* \nu+\mu \nu^*)e^{-2\alpha^2}$ but will be omitted below for simplicity.
While the unknown complex coefficients of the superposition $\mu,\nu\in \mathbb{C}$ are carrying encoded information, the amplitude $\alpha\in \mathbb{R}$ is typically assumed to be known. 
We focus primarily on nearly orthogonal coherence states with large $\alpha\gg2$ when this state can be considered as a qubit of basis $\ket{\pm \alpha}$.

Before going to the details of our protocol, let us first consider the decoherence of an unprotected 2-SCS  undergoing bosonic loss $\Gamma_l$.
Any density matrix $\rho$ under $\Gamma_l$ evolves as $\Gamma_l[\rho]=\sum_l \frac{(1-\eta)^l}{l!} \eta^{\hat{n}/2} \hat{a}^l \rho \hat{a}^{\dagger l} \eta^{\hat{n}/2}$ with $\eta$ the loss parameter.
2-SCS evolves by this loss simply as  
\begin{equation}
\Gamma_l[\ket{\psi_0}\bra{\psi_0}]=
|\mu|^2 \ket{\sqrt{\eta}\alpha}\bra{\sqrt{\eta}\alpha} + 
|\nu|^2 \ket{-\sqrt{\eta}\alpha}\bra{-\sqrt{\eta}\alpha} +
\mu\nu^*e^{-2\alpha^2(1-\eta)}\ket{\sqrt{\eta}\alpha}\bra{-\sqrt{\eta}\alpha} +
\mu^*\nu e^{-2\alpha^2(1-\eta)}\ket{-\sqrt{\eta}\alpha}\bra{\sqrt{\eta}\alpha}.
\label{eq:2scsloss}
\end{equation} 
 The off-diagonal elements carrying the quantum coherence in this nearly orthogonal coherent-state basis, are rapidly decreasing for a large $\alpha$ by the factor $e^{-2\alpha^2(1-\eta)}$.
This rapid decoherence can be avoided if $\alpha$ is effectively reduced, e.g. by mapping the  information  onto a two-level system.   


\begin{figure}[th]
\includegraphics[width=500px]{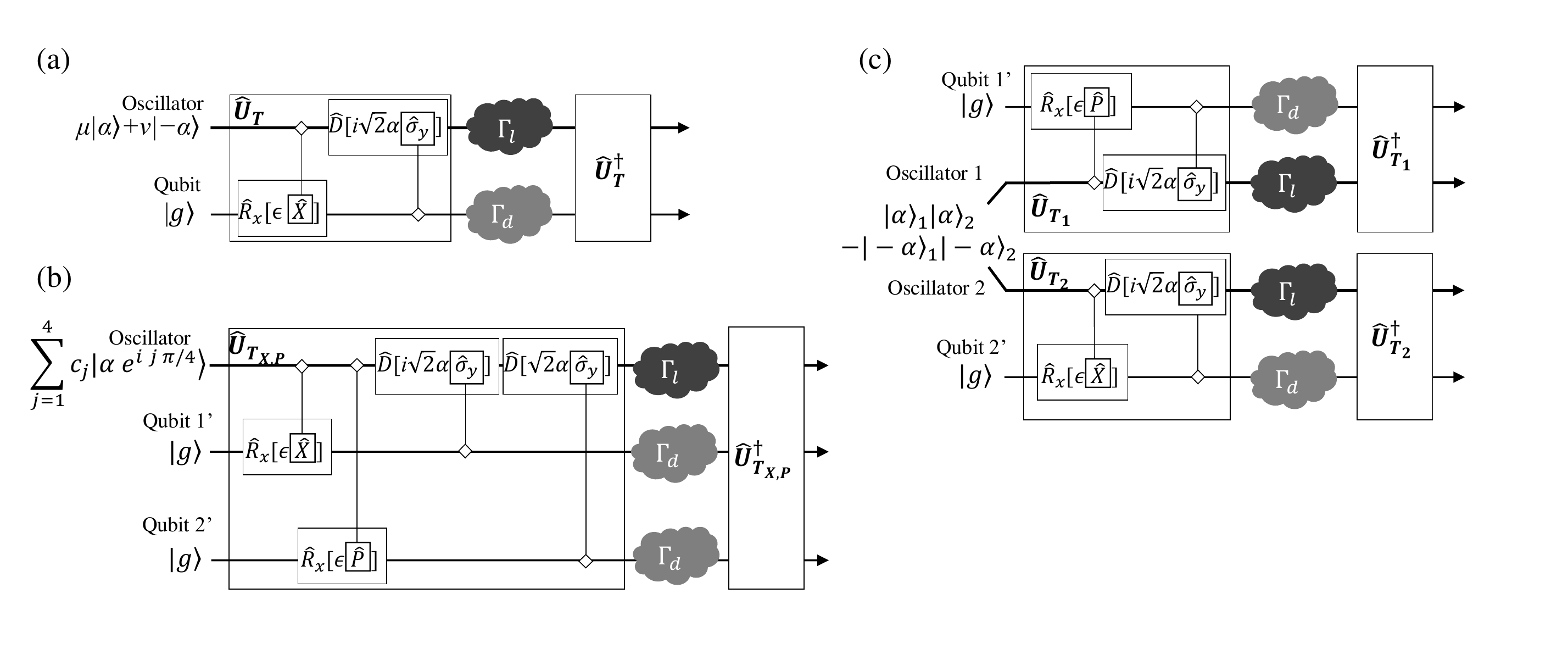} 
\caption{  (a) Schematic of the  protection protocol for a superposition of two coherent states in a bosonic mode (thick), exploiting a single qubit bypass (thin).
The oscillator and the ancillary qubit  interact by a unitary coupling $\hat{U}_T$ made of two Rabi gates, i.e. controlled qubit rotation $\hat{U}_R=\exp[i\epsilon \hat{\sigma}_x \hat{X}]=\hat{R}_x[\epsilon \hat{X}]$ where qubit rotation is denoted as $\hat{R}_x[\vartheta]=\exp[i\vartheta \hat{\sigma}_x]$, and controlled displacement $\hat{U}_D=\exp[-i \sqrt{2}\alpha \hat{\sigma}_y \hat{P}]=\hat{D}[i\sqrt{2}\alpha\hat{\sigma}_y]$. 
This bosonic channel  undergoes linear damping in a lossy channel, whereas lossless qubit  channel degrades only by a phase damping; the bosonic damping and phase damping are represented with trace-preserving maps $\Gamma_l$ and $\Gamma_d$, respectively. 
The bypass is intrinsically protected against the amplitude damping by the qubit encoding to two ground states. 
After $\hat{U}_T$, the quantum information is  transferred to the qubit substantially, but keeping the CV channel is still beneficial.  
After decoherence by the error channels, the input state is restored by the inverse unitary process $\hat{U}_T^{\dagger}$. 
An extension of the protocol exploiting two bypass qubits  (b) to a single-mode superposition of four coherent states, and
(c) to bipartite entangled states where the protocol is applied locally on each mode.
  } \label{fig:concept}
\end{figure}

Our bypass protocol illustrated in Fig.~\ref{fig:concept} (a) consists of a set of two Rabi gates 
$\hat{U}_T=\hat{U}_D \hat{U}_R$ applied before the lossy channel. 
The first unitary Rabi gate, $\hat{U}_R=\exp[i\epsilon \hat{\sigma}_x \hat{X}]$ with $\epsilon=\pi/4\sqrt{2}\alpha$, rotates the initial ancillary qubit state $\ket{g}$ controlled by the amplitudes in the coherent states to have an entangled state as:
\begin{align}
&\exp[i\epsilon \hat{\sigma}_x \hat{X}]\ket{g}(\mu\ket{\alpha}+\nu \ket{-\alpha})
=\mu \hat{D}[\alpha]\exp[i\epsilon \hat{\sigma}_x (\hat{X}+\sqrt{2}\alpha)]\ket{g}\ket{0}+\nu \hat{D}[-\alpha]\exp[i\epsilon \hat{\sigma}_x (\hat{X}-\sqrt{2}\alpha)]\ket{g}\ket{0}\nonumber\\
&=\frac{i}{2}\left\{\mu \hat{D}[\alpha]\left(\ket{-_i}\ket{\frac{i\epsilon}{\sqrt{2}}_+}-i \ket{+_i}\ket{\frac{i\epsilon}{\sqrt{2}}_-}\right)-\nu \hat{D}[-\alpha]\left(\ket{+_i}\ket{\frac{i\epsilon}{\sqrt{2}}_+}+i  \ket{-_i}\ket{\frac{i\epsilon}{\sqrt{2}}_-}\right)\right\}\nonumber\\
&\stackrel{\alpha\gg 2}{\longrightarrow}  i\left(\mu \ket{-_i}\ket{\alpha}-\nu \ket{+_i}\ket{-\alpha}\right),
\label{eq:psi1}
\end{align}
where an ordering relation $\hat{D}[\alpha]\hat{X} \hat{D}[-\alpha]=\hat{X}+\sqrt{2}\alpha$ was used on the first equality.
The approximation used on the last line is $\ket{\beta+i \varepsilon}\approx\exp[i\beta\varepsilon]\ket{\beta}$, for $\varepsilon\ll 1$.
Here, the unnormalized coherent superposition states are denoted as $\ket{\delta_\pm}=\ket{\delta}\pm \ket{-\delta}$ and
the qubit states $\ket{\pm_i}$ denote the eigenstates of $\hat{\sigma}_y$ with eigenvalues $\pm 1$. In the limit of a large amplitude  $\alpha\gg 2$~\cite{Bayer2017, Yoshihara2017},  the odd superpositions states vanish $\ket{\frac{i\epsilon}{\sqrt{2}}_-}\rightarrow 0$ and the even superpositions approach the vacuum  $\ket{\frac{i\epsilon}{\sqrt{2}}_+}\rightarrow \ket{0}$. 
In total, eq. (\ref{eq:psi1}) can be interpreted as entangling process of the 2-SCS and the qubit ancilla. 

The second Rabi gate $\hat{U}_D=\exp[-i \sqrt{2}\alpha \hat{\sigma}_y \hat{P}]=\hat{D}[i\sqrt{2}\alpha \hat{\sigma}_y ]$ acts as a controlled displacement, i.e. the  ancilla qubit encoded in $\ket{\pm_i}$ from (\ref{eq:psi1})  as the control and oscillator as the target,  which transfers the information to the qubit ancilla. 
Applying $\hat{U}_D$ on (\ref{eq:psi1}), we obtain the state which will undergo the noise channels:
\begin{align}
&\exp\left[-i\sqrt{2}\alpha\hat{\sigma}_y\hat{P}\right]\exp[i\epsilon \hat{\sigma}_x \hat{X}]\ket{g}(\mu\ket{\alpha}+\nu \ket{-\alpha})\nonumber\\
&=i\mu\left(\ket{-_i}\frac{\ket{i\epsilon/\sqrt{2}}+\ket{-i\epsilon/\sqrt{2}}}{2}+\ket{+_i}\frac{e^{-i\pi/4}\ket{2\alpha+i\epsilon/\sqrt{2}}-e^{i\pi/4}\ket{2\alpha-i\epsilon/\sqrt{2}}}{2i}\right)\nonumber\\
&-i\nu\left(\ket{+_i}\frac{\ket{i\epsilon/\sqrt{2}}+\ket{-i\epsilon/\sqrt{2}}}{2}-\ket{-_i}\frac{e^{i\pi/4}\ket{-2\alpha+i\epsilon/\sqrt{2}}-e^{-i\pi/4}\ket{-2\alpha-i\epsilon/\sqrt{2}}}{2i}\right)\nonumber\\
&\stackrel{\alpha\gg 2}{\longrightarrow} i\left(\mu\ket{-_i} -\nu\ket{+_i}\right)\ket{0}.
\label{eq:input}
\end{align}
This state contains even superposition of coherent states  $\ket{\pm i\epsilon/\sqrt{2}}$ close to a vacuum state and erroneous states  $\ket{\pm 2\alpha\pm i\epsilon/\sqrt{2}}$ far from the phase space origin.
We note that only the erroneous terms are affected heavily by loss due to their large amplitudes, and it can be still advantageous to keep the bosonic channel. 




We consider qubit ancilla to be lossless without amplitude damping, but sensitive to small phase damping~\cite{Bruzewicz2019APRtrappedion,Kjaergaard2020SuperCondQub}  as the main threat to robustness. We model the phase damping  as $\Gamma_d[\rho]=(1-\tfrac{p}{2})\rho+\tfrac{p}{2} \hat{\sigma}_z \rho \hat{\sigma}_z$ for any state $\rho$ with dephasing parameter $p$. 
This channel is identity channel at $p=0$, and maximal dephasing channel at $p=1$.
The same dephasing channels  act independently on all physical qubit ancillas involved through the paper.
Note that it is also  possible to correct this dephasing error actively by transferring the information into multiple ancillas for conventional DV error correction methods~\cite{LidatQEDBook2013, Gaitan2008}, or by  a hybrid qubit protection scheme using the Rabi type of interactions as in the Appendix \ref{sec:AppB}. 
In section \ref{sec:results}, we show that after the error channels $\Gamma_l$ and $\Gamma_d$, the original state is substantially restored by the inverse operations $\hat{U}_T^{\dagger}$.
We also note that in principle, the entire protocol can be achieved by dispersive interaction and displacement \cite{SpillerNJP2006,vanLoock2008HybridOptics} in travelling-wave microwave/optical platforms for future  experiments extending the scheme in \cite{HackerNatPho2019}.


{\em Four-coherent state superposition---} This methodology can be scaled up to cases of more coherent state components in the superposition, e.g. to 4-SCS $\mu_1\ket{\alpha }+\mu_2\ket{\alpha ^*}+\mu_3\ket{-\alpha }+\mu_4\ket{-\alpha^*}$ for $\alpha=\alpha_r+i\alpha_i\in\mathbb{C}$~\cite{Hastrup20194hc}. 
Gaussian operations such as squeezing cannot protect such states \emph{at all}, as the broken symmetry in phase space by squeezing leads to a more severe decoherence. 
Therefore, these states  can be an advanced test bed for the extension of the bypass strategy to general SCSs. 
We note that a single qubit will not be sufficient for the bypass due to the existence of 4 unknown coefficients $\mu_j$'s, and we need to extend the circuit to two ancilla qubits for the  minimal complexity. 


 To protect a general 4-SCS with our hybrid bypass scheme,  two ancillas with indices $j=1',2'$  couple  respectively to the quadrature variables $X$ and $P$ of oscillator  by four Rabi gates forming $\hat{U}_{T_{X,P}}$ as depicted in Fig.~\ref{fig:concept} (b).  
 The two quadratures are independently coupled to each of the ancillas  by the operations $\hat{U}_R^{(1')}=\exp[i\epsilon_r \hat{\sigma}_x^{(1')} \hat{X}]$ and $\hat{U}_R^{(2')}=\exp[i\epsilon_i \hat{\sigma}_x^{(2')} \hat{P}]$ with a single Pauli operator acting on  ancilla $j$ $\hat{\sigma}_x^{(j)}$,  approximated in the large amplitude limit $\alpha_{r,i}\gg 2$ as:
\begin{align}
&\exp[i\epsilon_r \hat{\sigma}_x^{(1')} \hat{X}]\ket{g}_{1'}\ket{\alpha_r+ i \alpha_i}
\approx (\cos[\sqrt{2}\alpha_r \epsilon_r]\ket{g}_{1'}+ i\sin[\sqrt{2}\alpha_r \epsilon_r]\ket{g'}_{1'})\ket{\alpha_r+ i \alpha_i}\approx  \ket{\mathrm{sgn}[\alpha_r]_i}_{1'}\ket{\alpha_r+ i \alpha_i}, \nonumber\\
&\exp[i\epsilon_i \hat{\sigma}_x^{(2')} \hat{P}]\ket{g}_{2'}\ket{\alpha_r+ i \alpha_i}\approx (\cos[\sqrt{2}\alpha_i \epsilon_i]\ket{g}_{2'}+ i\sin[\sqrt{2}\alpha_i \epsilon_i]\ket{g'}_{2'})\ket{\alpha_r+ i \alpha_i}\approx\ket{\mathrm{sgn}[\alpha_i]_i}_{2'}\ket{\alpha_r+ i \alpha_i}.
\end{align}
Here, $\mathrm{sgn}[x]$ denotes the sign of the argument $x$. 
Coupling strengths are set respectively as $\epsilon_{r,i}=\pi/4\sqrt{2}\alpha_{r,i}$.
The coherent peaks are then shifted toward the phase space origin by the controlled displacement $\hat{U}_D^{(j)}$  by the ancillary states before the erroneous channels as
\begin{align}
&\exp[i \sqrt{2}\alpha_r \hat{\sigma}_y^{(1')} \hat{P}]\ket{\mathrm{sgn}[\alpha_r]_i}_{1'}\ket{\alpha_r+ i \alpha_i}\approx \ket{\mathrm{sgn}[\alpha_r]_i}_{1'}\ket{i\alpha_i}, \nonumber\\
&\exp[-i \sqrt{2}\alpha_i \hat{\sigma}_y^{(2')} \hat{X}]\ket{\mathrm{sgn}[\alpha_i]_i}_{2'}\ket{i \alpha_i}\approx \ket{\mathrm{sgn}[\alpha_i]_i}_{2'}\ket{0}.
\end{align}
 After the error channels, the original state can be substantially restored by the inverse operations $\hat{U}_{T_{X,P}}^{ \dagger}$.
This example demonstrates that complex n-SCSs can be protected by the qubit bypass in similar ways; for other examples see Appendix F.

{\em Two-mode entangled coherent states---}Our scheme  can be extended to multi-mode ECSs in a straightforward way, for the simplest example to the two-mode state $\ket{\mathrm{ECS}^{\pm}}=N_\pm(\ket{\alpha}_1\ket{\alpha}_2\pm\ket{-\alpha}_1\ket{-\alpha}_2)$ with oscillator indices $J=1,2$ with a normalization factor $N_\pm=(2\pm 2e^{-4\alpha^2})^{-1/2}$. 
Without any protective scheme, the two-mode ECS will decohere at approximately double decay rate of single mode superposition states as 
\begin{align}
\Gamma_l^{(2)}[\Gamma_l^{(1)}[\ket{\mathrm{ECS}^{\pm}}\bra{\mathrm{ECS}^{\pm}}]]
=&N_\pm^2( \ket{\alpha'}_1\bra{\alpha'}\otimes \ket{\alpha'}_2\bra{\alpha'}+\ket{-\alpha'}_1\bra{-\alpha'}\otimes \ket{-\alpha'}_2\bra{-\alpha'}\nonumber\\
&\pm e^{-4\alpha^2(1-\eta)}\ket{\alpha'}_1\bra{-\alpha'}\otimes \ket{\alpha'}_2\bra{-\alpha'}\pm e^{-4\alpha^2(1-\eta)}\ket{-\alpha'}_1\bra{\alpha'}\otimes \ket{-\alpha'}_2\bra{\alpha'}),
\end{align}
where $\alpha'=\sqrt{\eta}\alpha$.
For this state, our scheme utilizes two \emph{local} bypasses  of ancillary modes $1',2'$ acting on both CV modes. 
The state  $\ket{\mathrm{ECS}^{\pm}}$ then evolves by $\hat{U}_{T_{1}}\hat{U}_{T_{2}}$ to a loss-robust state
\begin{align}
&\hat{U}_{T_{1}}\hat{U}_{T_{2}}\ket{\mathrm{ECS}^{\pm}}=i\ket{\phi_+}_{1,1'}\ket{\phi_+}_{2,2'}\mp i\ket{\phi_-}_{1,1'}\ket{\phi_-}_{2,2'} \label{eq:4hcdm}
\end{align}
where $\ket{\phi_+}=\ket{-_i}\frac{\ket{i\epsilon/\sqrt{2}}+\ket{-i\epsilon/\sqrt{2}}}{2}+\ket{+_i}\frac{e^{-i\pi/4}\ket{2\alpha+i\epsilon/\sqrt{2}}-e^{i\pi/4}\ket{2\alpha-i\epsilon/\sqrt{2}}}{2i}$
and $\ket{\phi_-}=\ket{+_i}\frac{\ket{i\epsilon/\sqrt{2}}+\ket{-i\epsilon/\sqrt{2}}}{2}-\ket{-_i}\frac{e^{i\pi/4}\ket{-2\alpha+i\epsilon/\sqrt{2}}-e^{-i\pi/4}\ket{-2\alpha-i\epsilon/\sqrt{2}}}{2i}$.
 For an $n$-mode ECSs, we can similarly apply local schemes to all $n$ modes to transform into more loss-robust states.
These examples explain well the logical steps to be applied to any specific superposition of finite number of coherent states in different modes.   



\section{Results}
\label{sec:results}

\subsection{Single mode superposition states}

As a measure of performance of the protocol, we can first look at fidelity \cite{Jozsa1994Fidelity, Uhlmann1976Fidelity}. 
Fidelity is defined as $F(\rho_\mathrm{in},\rho_\mathrm{out})=\mathrm{tr}(\sqrt{\sqrt{\rho_\mathrm{in}}\rho_\mathrm{out}\sqrt{\rho_\mathrm{in}}})^2$ between input $\rho_\mathrm{in}$ and output $\rho_\mathrm{out}$.
For large $\alpha\gg2$, the superpositions of coherent states represent a qubit space that is well characterized by fidelity. 
The average performance of the schemes on 2-SCSs $\mu\ket{\alpha}+\nu\ket{-\alpha}$ with a fixed $\alpha$ but arbitrary $\mu$ and $\nu$ over the ``Bloch sphere" of coherent states can be measured by the  channel fidelity~\cite{AlbertPRA2018bosoniccode}. 
It is defined as the fidelity between an input virtual entangled state $2^{-1/2}(\ket{\mathbf{0}}\ket{\alpha}+\ket{\mathbf{1}}\ket{-\alpha})$ and the output state after the noise channels, where $\{\ket{\mathbf{0}},\ket{\mathbf{1}}\}$ are fictitious qubit basis. 
We note that in contrast to a conventional qubit, the statistical weight of states represented in the channel fidelity is not homogeneous over all coefficients $\mu, \nu$.
For example,  the weight of balanced superposition states $\ket{\alpha}\pm \ket{-\alpha}$ has the extrema values, but this inhomogeneity is negligible for $\alpha\gtrsim2$.  

\begin{figure}[th]
\includegraphics[width=450px]{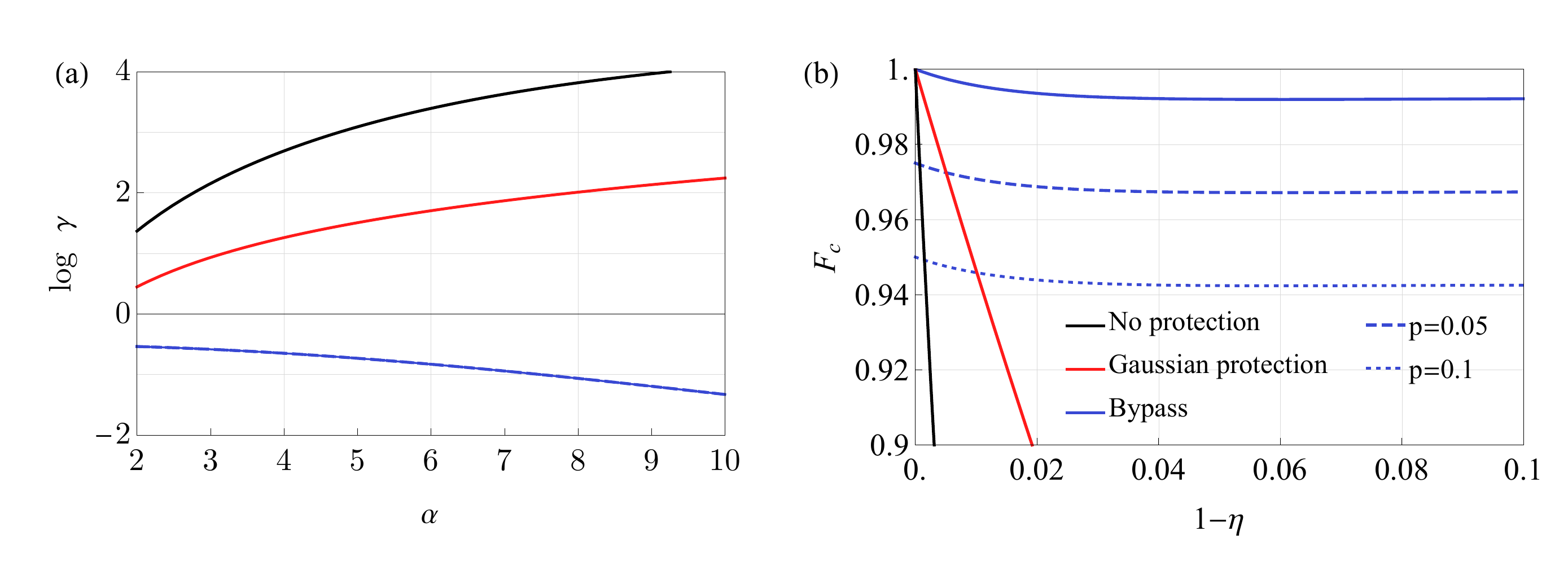}
\caption{ Performance of bypass protection protocol in Fig.~\ref{fig:concept} (a) for a superposition of two coherent states in a bosonic mode compared to the Gaussian strategy under various dephasings. 
 (a) The logarithmic decay rate $\log \gamma$ of channel fidelity modeled as $F_c(\eta)=F_c(1)e^{-\gamma(1-\eta)}$ for various cases.
 The curves for different $p$'s are overlapped.
(b) Channel fidelity vs. loss parameter for $\alpha=6$. 
This result shows the qualitative difference between bypass and Gaussian strategies.
 } \label{fig:2hcL}
\end{figure}

We compared the strategy of Gaussian squeezing strategy prior to transmission through the lossy channel~\cite{FilipPRA2013GaussianAdaptation,
LeJeannicPRL2018} as a  benchmark.
The optimization of Gaussian strategy was with respect to the average photon number in the CV modes at each $\alpha$.
The optimal squeezing parameter $r_\mathrm{opt}$ for the initial state $\mu\ket{\alpha}+\nu\ket{-\alpha}$ is given as
\begin{align}
r_\mathrm{opt}=\frac{1}{4} \left(\log \left[\frac{4 e^{2 \alpha ^2} \alpha ^2 \mu ^*}{e^{2 \alpha ^2} \mu ^*+\nu  \left(\mu
   ^*\right)^2+\mu ^2 \nu }+1\right]-\log \left[1-\frac{4 \alpha ^2 \nu  \left(\left(\mu ^*\right)^2+\mu ^2\right)}{e^{2
   \alpha ^2} \mu ^*+\nu  \left(\mu ^*\right)^2+\mu ^2 \nu }\right]\right).
\end{align}

In Fig.~\ref{fig:2hcL}, we show the effectiveness of our strategy quantified by the channel fidelity for general 2-SCSs of various amplitudes against bosonic loss   (fidelity for individual states are calculated in Appendix~\ref{sec:individ}). 
This is well evidenced by the decay rate $\gamma$ of the fidelity model $F_c(\eta)=F_C(1)e^{-\gamma(1-\eta)}$, where a larger $\gamma$ represents a faster decay.
Fig.~\ref{fig:2hcL} (a) clearly shows the qualitative difference of our qubit bypass strategy showing decreased $\gamma$ approximately described as $0.65- 0.025 \alpha^{1.2}$.
In contrast, the $\gamma$s for no protection and Gaussian squeezing protection are described by increasing trends respectively as $0.17+\alpha^{1.92}$ and  $0.29+0.96\alpha$. 
In Fig.~\ref{fig:2hcL} (b), we can see that the channel fidelity of  bypass scheme  significantly surpasses the Gaussian strategy for a mild level of dephasing for a large amplitude $\alpha\gg 2$, and  loss is not critically detrimental to our scheme. 
At complete lossy case $\eta=0$ for $p=0$ corresponding to the complete blockage of the oscillator, the  channel fidelity of our scheme can be approximately given in the large $\alpha$ limit as
\begin{align}
F_c\approx\frac{e^{-\frac{3 \pi ^2}{64 \alpha ^2}}+e^{-\frac{\pi ^2}{64 \alpha ^2}}}{2}.
\label{eq:asymptfid}
\end{align}
We can see in Fig.~\ref{fig:2hcFCa} (a,b) in Appendix \ref{sec:individ} that keeping the oscillator is still beneficial even though it carries little information.

\begin{figure}[th]
\includegraphics[width=450px]{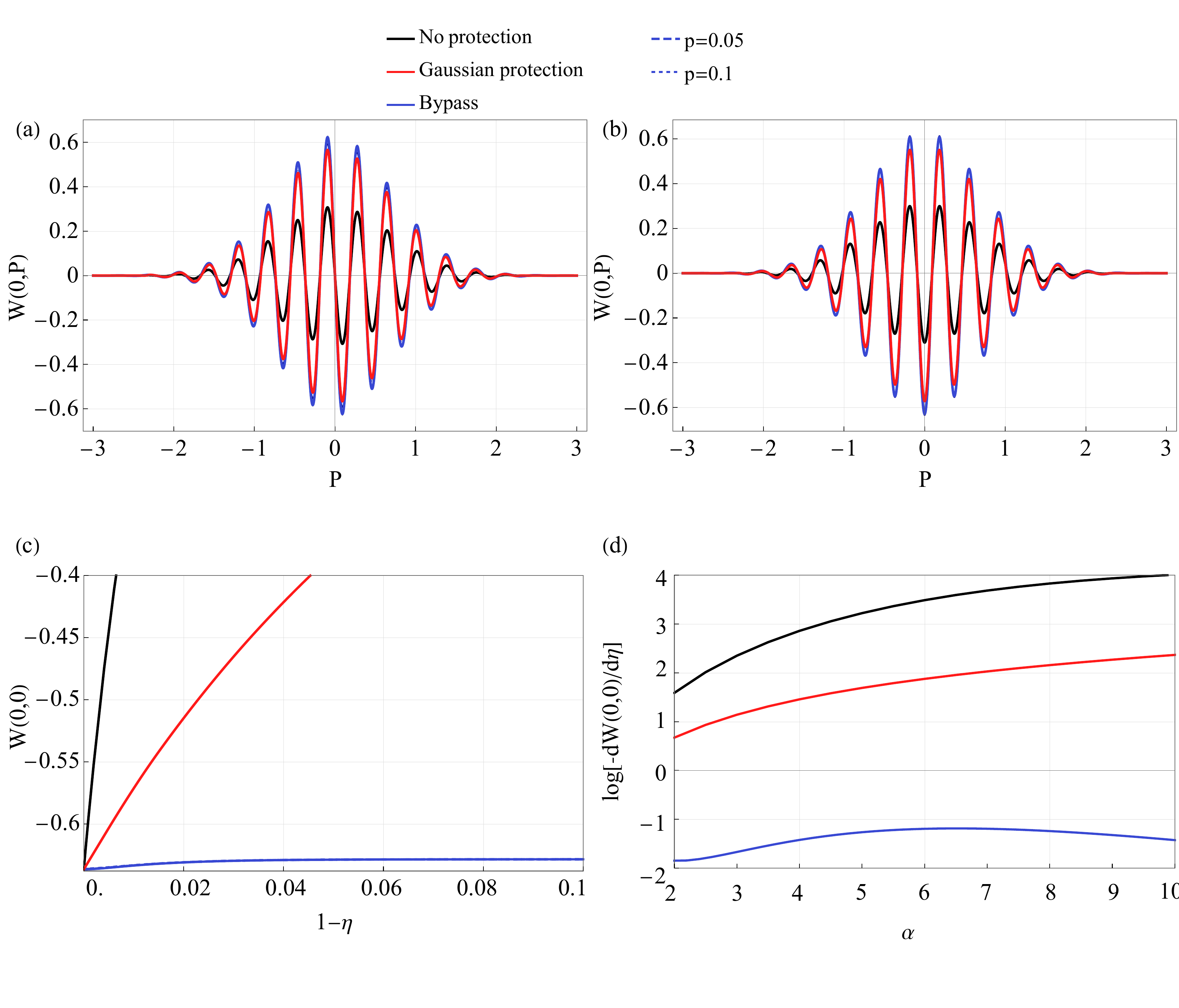}
\caption{
Protection of negative peaks of Wigner functions of superposition of two coherent states as in Fig.~\ref{fig:concept}~(b).
(a,b)  Interference fringes in cross section of the Wigner function  for $\ket{\alpha}+e^{i\phi}\ket{-\alpha}$ at $\alpha=6$ with $\phi=\pi/2,\pi$ under $1-\eta=0.01$ of loss and $p=0,0.05,0.1$ of dephasing. 
The interference fringes for complete lossy channel $\eta=0$ and under various $p$'s are overlapped largely with the initial curve.
(c) The  depths of the negative peak located at phase space origin vs. loss parameter $1-\eta$ at $\alpha=6$ for the initial state $\ket{\alpha}-\ket{-\alpha}$ (curves with $p>0$ are overlapped), (d) and the logarithm  of  slopes of the curves in (c) vs. $\alpha$.
} \label{fig:intfri}
\end{figure}

The negative part of the Wigner functions can serve as a quantitative measure of the  non-classical features, highly sensitive to both losses and noises~\cite{ZurekRMP2003Decoherence, SchlosshauerRMP2005Decoherence}. 
The equal-weight 2-SCSs $\ket{\alpha}+e^{i\phi}\ket{-\alpha}$ have interference fringes around the origin of the phase space.
In Fig.~\ref{fig:intfri} (a), these interference fringes  along P-axis in phase space are shown  for various $\phi$ under various levels of loss and dephasing. 
We note that the negativity of all interference fringes are preserved, not only in the largest one. 
The bypass scheme protects the interference fringes even under a large dephasing, especially for a large $\alpha\gg 2$.
Although the input state $\ket{\alpha}+i\ket{-\alpha}$ is not fully protected  in terms of fidelity under dephasing,  the interference fringes are not critically destroyed, although not immune.
Mathematically, the value of the Wigner function at the phase space origin is proportional to the average value of the parity operator as $W(0,0)=\frac{2}{\pi}\langle\rho  {(-1)}^{\hat{n}}\rangle$.
The bypass scheme protects the interference fringes even under a large dephasing, especially for a large $\alpha\gg 2$.
The deepest negative peak of the initial state $\ket{\alpha}-\ket{-\alpha}$ is located at the phase space origin, and its first-order approximation for our protocol vs. the loss parameter $1-\eta$ and  the dephasing parameter $p$  scales approximately in the large $\alpha$ limit as
\begin{align}
W(0,0)\approx-\frac{2-4 e^{-2 \alpha ^2}}{\pi }+(1-\eta)\frac{2 e^{-2 \alpha ^2} \alpha ^2}{\pi }+p \frac{1-e^{-\frac{\pi ^2}{8 \alpha ^2}}}{2 \pi }.
\end{align}
Here the effect of  loss and dephasing evidenced in the second and third terms is \emph{weakening} for a large $\alpha$.
This is because the state $\ket{\alpha}-\ket{-\alpha}$ is mapped onto the oscillator ground state (Eq.~(\ref{eq:input})), which is unaffected by both boson loss and dephasing. 
Fig.~\ref{fig:intfri} (b) shows that our scheme protects this negative peak better than the Gaussian squeezing protection especially at large $\alpha$, whose negative peak at the origin is described approximately by $-\frac{2}{\pi}+0.61 (1-\eta)  \alpha^{1.60}$. 
In comparison, without any protection the negative peak at the origin scales as $\frac{2 e^{2 \alpha ^2 (1-\eta)}-2 e^{2 \alpha ^2 \eta }}{\pi  \left(e^{2
   \alpha ^2}-1\right)}$, which is influenced more harshly than the others.



In Appendix \ref{sec:4scs}, we calculated the Wigner functions and channel fidelity of the 4 SCSs for a special encodings as in~\cite{LeghtasPRL2013,OfekNature2016CatExperiment} under various levels of loss and dephasing. 
Similarly as for 2-SCSs, we can also protect the negative peaks of 4-SCSs  with an increased fidelity in large $\alpha$ limit against  loss  under a modest level of dephasing on the ancilla. 


\subsection{Two mode entangled states}

\begin{figure}[th]
\includegraphics[width=450px]{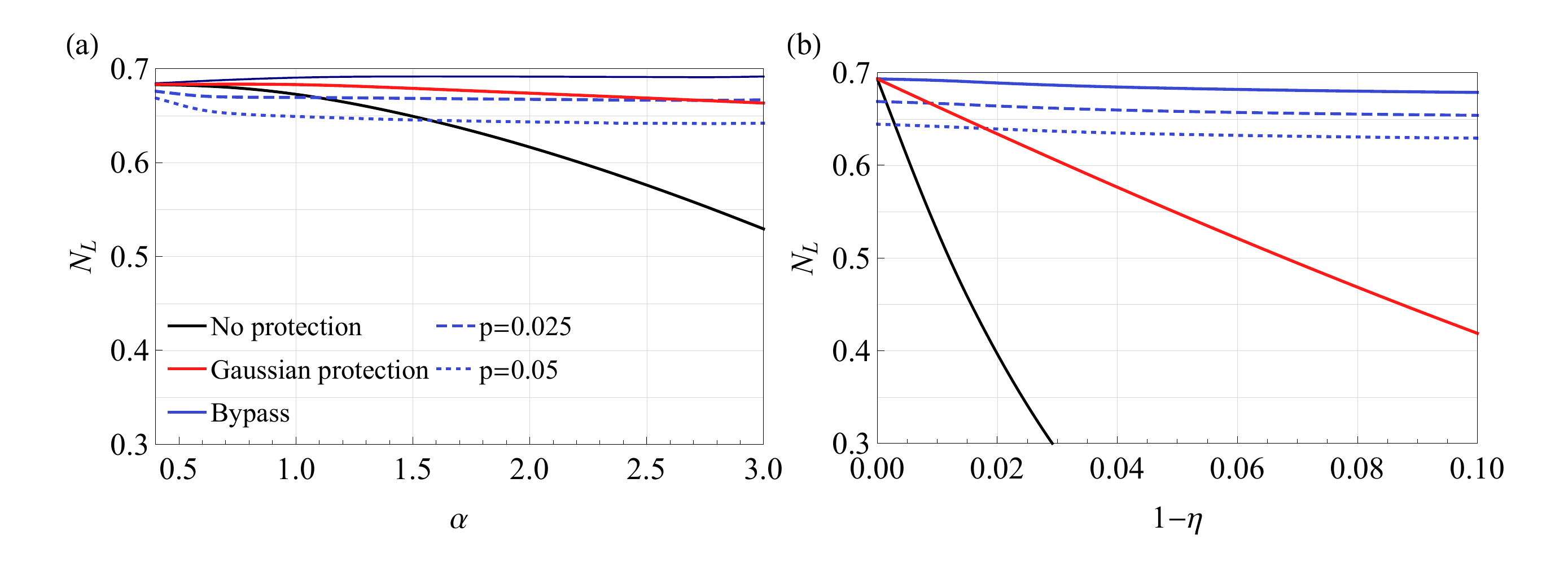}
\caption{ Protection of entangled coherent states $\ket{\mathrm{ECS}_-}$ in Fig.~\ref{fig:concept} (c) quantified by numerically calculated logarithmic negativities 
 (a) vs. $\alpha$ for $1-\eta=0.01$ and (b) vs. $1-\eta$ for $\alpha=3$.  } \label{fig:NegEcs}
\end{figure}


Entangled coherent states \cite{SandersRev2012ECS,Ourjoumtsev2009ECS} were engineered experimentally from trapped ions \cite{SarlettePRA2012TrappedIonECS, Arenz2013TrappedIonECS} and superconducting circuits \cite{WangScience2016ECS,WECSgeneration}. 
We briefly note that in terms of fidelity, similar tendencies, e.g. enhancement at large $\alpha$ beyond Gaussian strategies,  can be obtained as of single mode states.
An important question is whether a non-local property such as  entanglement of these states can also be protected by the local bypass schemes as in Fig.~\ref{fig:concept} (c), which cannot be simply predicted by the fidelity.
For our protocol, intuitively we expect this to be the case, as we transfer the entanglement from the bosonic modes to the ancillas, thus bypassing losses.
As a measure of entanglement which can characterize the bipartite states $\rho$, we choose logarithmic negativity $N_L[\rho]=\log||\rho^\mathrm{PT}||$ with partial transposition $\mathrm{PT}$ and the trace norm $||\cdot||$, which operationally is connected to the upper bound on the distillable entanglement~\cite{PlenioPRL2005Logneg}.   
The even superposition state $\ket{\mathrm{ECS}^{+}}$ has $0$ logarithmic negativity at small $\alpha$  as it is close to the vacuum and thus has no entanglement. 
On the other hand, the odd superposition state $\ket{\mathrm{ECS}^{-}}$  possesses a value $\ln 2$ at all $\alpha$.

At each $\alpha$, the squeezing parameters for Gaussian strategies were optimized to minimize the average photon number before the lossy channel.
The Gaussian squeezing protection can be applied on both modes, with the optimal squeezing parameter with respect to average photon number is given as
\begin{align}
r_\mathrm{opt}=\frac{1}{4} \left(\log \left[2 \alpha ^2+2 \alpha ^2 \coth \left(2 \alpha ^2\right)+1\right]-\log \left[-2 \alpha ^2+2   \alpha ^2 \coth \left(2 \alpha ^2\right)+1\right]\right).
\end{align}
In addition, we optimized the Gaussian protection numerically at each $\alpha$ as well.

The scaling of the logarithmic negativity vs. $\alpha$ and $\eta$ can be described by a fitted function $\log 2-(1-\eta)f[\alpha]$ when the dephasing is absent.
The  loss effect functions are approximately given as $f[\alpha]=1.59 \alpha ^2+0.12 \alpha +0.59$ for no protection,
$f[\alpha]=-0.26+1.34 \alpha -0.70 \log \alpha $ for Gaussian squeezing protection, and $f[\alpha]=0.44 -0.22 \log\alpha$ for bypass protocol.
In Fig.~\ref{fig:NegEcs},  the curves for the logarithmic negativities of ECS $\ket{\mathrm{ECS}^{-}}$ after various noise channels were drawn numerically.
We notice again that   the entanglement of the bipartite states suffers weaker effect
by qubit bypass protocol
than the Gaussian strategies, with enhanced performance in large $\alpha$ limit  under a moderate qubit dephasing. 

\begin{figure}[th]
\includegraphics[width=450px]{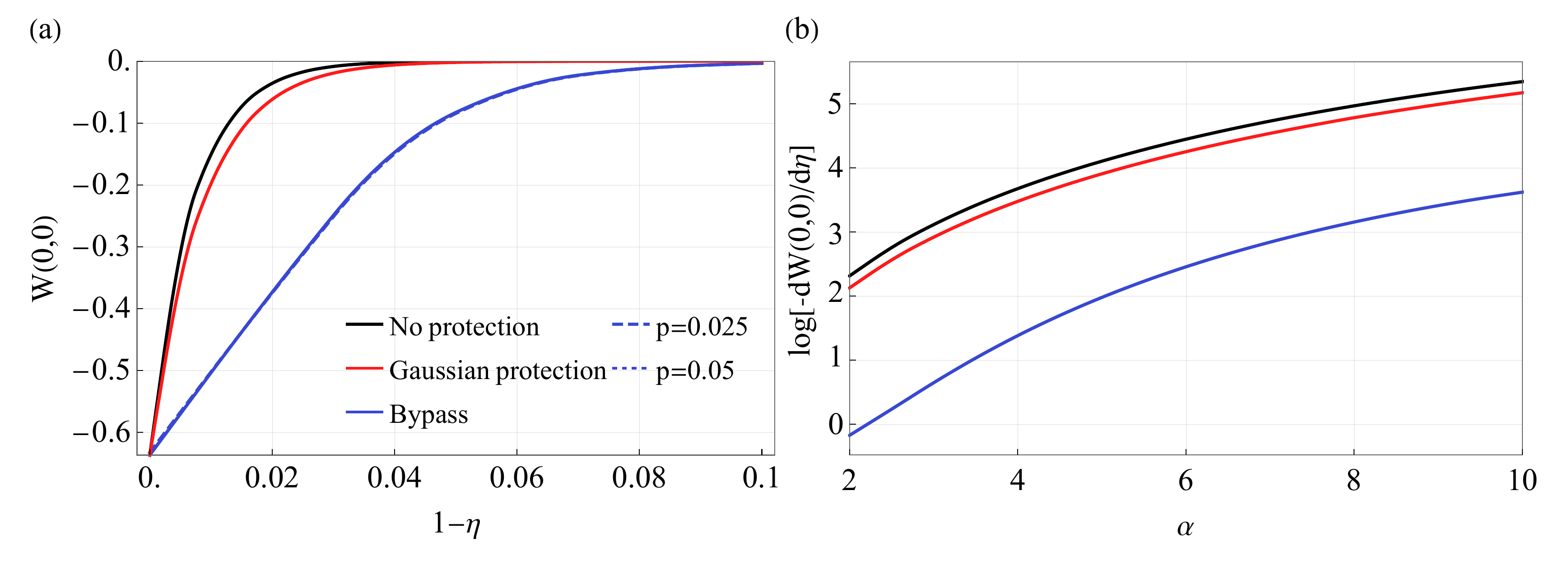}
\caption{ Protection in Fig.~\ref{fig:concept}~(c) of the Wigner function negative peak depths at the phase space origin of $\ket{\mathrm{ECS}^-}$ after a projective vacuum measurement on one mode. (a) vs. loss parameter $1-\eta$ at $\alpha=6$ (curves for $p>0$ are overlapped), (b) the logarithm of the slope of the curves in (a). 
The curves for the protection by squeezing were calculated numerically.
The dephasing noise does not affect the depth of the negative peaks, as is evidenced by the overlap of the curves.
} \label{fig:steering}
\end{figure}

Additionally, the negative Wigner function of the projected state by a local detection
shows a sensitive witness of the entanglement. 
If the bipartite state becomes factorizable in coherent basis due to decoherence as  $\ket{\alpha}_1\bra{\alpha}\otimes \ket{\alpha}_2\bra{\alpha}+\ket{-\alpha}_1\bra{-\alpha}\otimes \ket{-\alpha}_2\bra{-\alpha}$,  no measurement in mode $1$ can create a negative peak in the phase space of mode $2$.
In contrast, ECSs allow measurement-induced preparation of coherent superposition states which possess negative Wigner function. 
As the simplest examples, we can consider  homodyne measurements $\ket{P}_1\bra{P}$ or $\ket{X}_1\bra{X}$ with $X=0$ or $P=0$, or a projection on vacuum state $\ket{0}_1\bra{0}$.
For an ideal entangled state $\ket{\mathrm{ECS}^{\pm}}$, all projections will prepare superposition states $\ket{\alpha}_2\pm\ket{-\alpha}_2$.  
In Fig.~\ref{fig:steering}, we compared how the deepest negative peak  of Wigner function behaves for different $\alpha$ and $\eta$.
Without protection, the formula for the largest negative peak depth is given as $w_0=\frac{2 \left(e^{-4 \alpha ^2 (\eta -1)}-e^{2 \alpha ^2 \eta
   }\right)}{\pi  \left(e^{-2 \alpha ^2 (\eta -2)}-1\right)}$, and by squeezing protection approximately as $(w_0)^{e^{-0.18}}$.
The approximate formula for qubit bypass scheme is given as $-\frac{2 \exp[{\left(1.46 \alpha ^2-7.14 \alpha +11.0\right) (\eta -1)}}{\pi }]$.
We can see that our protocol is clearly superior to the Gaussian squeezing strategies in protecting the locally induced negativity of of Wigner functions in all parameter regions of $\alpha$s and $\eta$s.
Interestingly, the squeezing is not highly effective in protecting the negative interference fringes here. 


Notably for a large $\alpha$, this Wigner function negative peak of the local projected state is impacted by loss even though the entanglement itself is preserved by our protection.
This discrepancy is caused by the effect of loss on local projection in the control mode.
It might limit some applicability of the protected states to non-local tasks in multi-modes, and further evaluation of the loss tolerance of conditional states may be needed. 
We speculate that certain non-local protocols may enhance  the protection of non-local aspects.





\section{Conclusions}
\label{sec:conclusion}

In this work, we proposed a deterministic  bypass protocol utiliaing Rabi couplings to lossless  qubit ancillas under a low level of dephasing, to mitigate decoherence by stable loss on the unknown single-mode and two-mode superposition of coherent states with various number of components.
 Our method depends on experimental realization of the Rabi couplings:  high strengths of the coupling, and the existence of counter-rotating term in the Hamiltonians.
The generation of high strength gate can be in principle  assisted by inline squeezings (see Appendix \ref{sec:JCsqueezing}).
This protocol can be implemented in various cavity QED systems which support the precise control of Rabi couplings~\cite{SolanoRabiRMP2019,NoriRabiRMP2019,MuellerNature2020Plasmon, Fluhmann2018Rabi, Lv2018RabiTrappedIon, LangfordNatComm2017Rabi, Markovic2018RabiSC,vanLoock2008HybridOptics}.
We analyzed the performance by various measures in comparison to optimal Gaussian strategies, and confirmed the superiority of our protocol.

Our protocol possesses following unique characteristics: (i) it utilizes a minimal number of  resources in contrast to ultimate but challenging quantum error correction schemes, (ii) the performance improves for larger $\alpha$ in contrast to existing protocols optimal in the opposite regime,  (iii) CV channel plays  non-negligible role except for a very large $\alpha\gtrsim 10$,
and (iv) it can be combined with conventional error correction schemes on qubits~\cite{OfekNature2016CatExperiment,Noh2019GKP, Fabre2019Combs,HeeresNatComm2017,Hastrup2019GKP,MichaelPRA2016Binomial, NohPRA2020SurfaceGKP, FluhmannHomeNature2019, SchoelkopfNature2020,GrimmNature2020KerrCat,HuNatPhys2019Binomial,PuriSciAdv2020Cat,TzitrinPRA2020GKP,TerhalQST2020GKP} or a Gaussian protection scheme on CV states~\cite{LeJeannicPRL2018,FilipJOB2001Amplifier,Serafini2004MinimumDecoherenceCat, FilipPRA2013GaussianAdaptation}. 
Our work implies that robust qubit channels can be exploited for future extension of the method to various classes of CV quantum non-Gaussian or non-local resources.









\appendix

\section*{Acknowledgment}
This project was supported by the Danish National Research
Foundation through the Center of Excellence for-
Macroscopic Quantum States (bigQ, DNRF0142). R.F. acknowledges
project LTAUSA19099 and 8C20002 from the Ministry
of Education, Youth and Sports of Czech Republic. K.P.
acknowledges project 19-19722J of the Grant Agency of
Czech Republic (GA\v{C}R).
This project has received funding from the European Union’s 2020
research and innovation programme (CSA - Coordination and support
action, H2020-WIDESPREAD-2020-5) under grant agreement No. 951737
(NONGAUSS).

\section*{Data availability}
The numerical data presented in this study is available from the authors upon request.


\section*{Author contributions}
All authors designed the protocol. K.P. and J.H. performed the calculations.  R.F. and U.L.A. supervised the work. All authors discussed and interpreted the results, and contributed to the writing of the manuscript.

\section*{Competing interest}
The authors declare that there are no competing interests.
\appendix
\counterwithin{figure}{section}
\section{Channel fidelity vs. $\alpha$ and fidelity for individual 2-SCS}
\label{sec:individ}

\begin{figure}[th]
\includegraphics[width=450px]{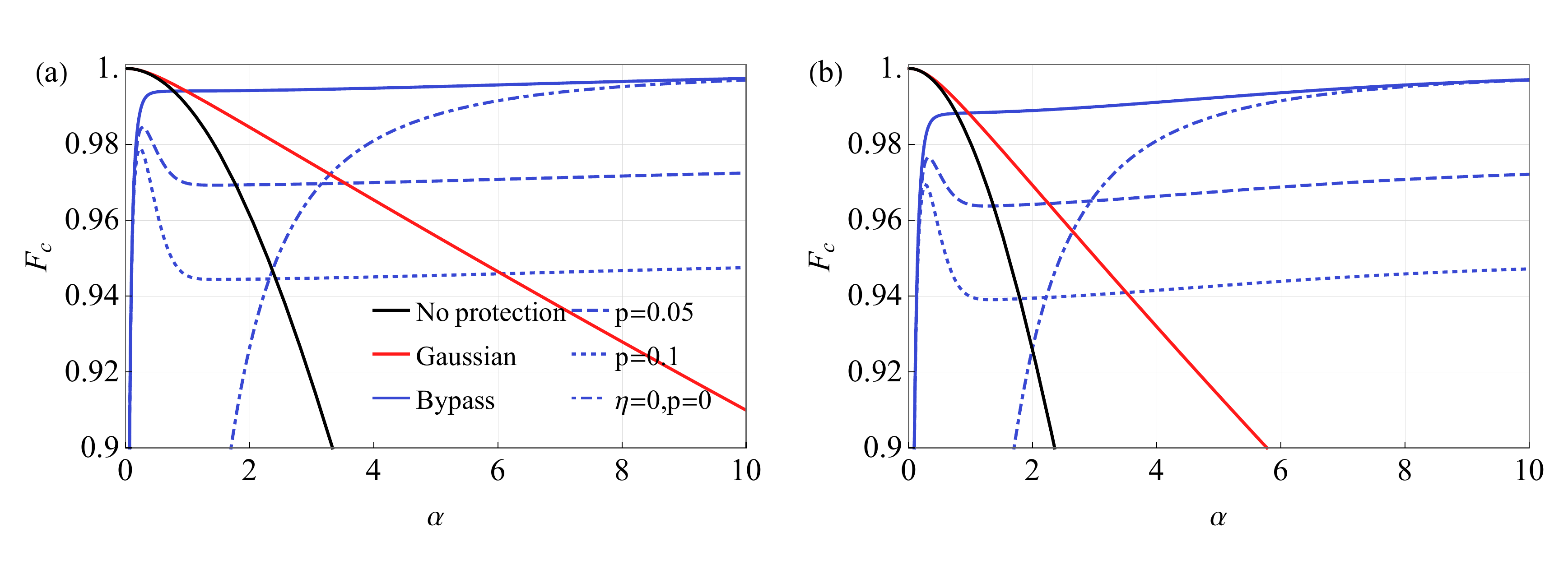}
\caption{Channel fidelity vs. coherent amplitude $\alpha$ under losses (a) $1-\eta=0.01$, (b) $0.02$,  and dephasings $p=0.00,0.05,0.10$. } \label{fig:2hcFCa}
\end{figure}

In Fig.~\ref{fig:2hcFCa}, the channel fidelity vs. $\alpha$ are shown at  small losses.
We can observe a clear enhancement over Gaussian strategy.
We can also see    that keeping the oscillator is still beneficial even though it carries little information.

\begin{figure}[th]
\includegraphics[width=500px]{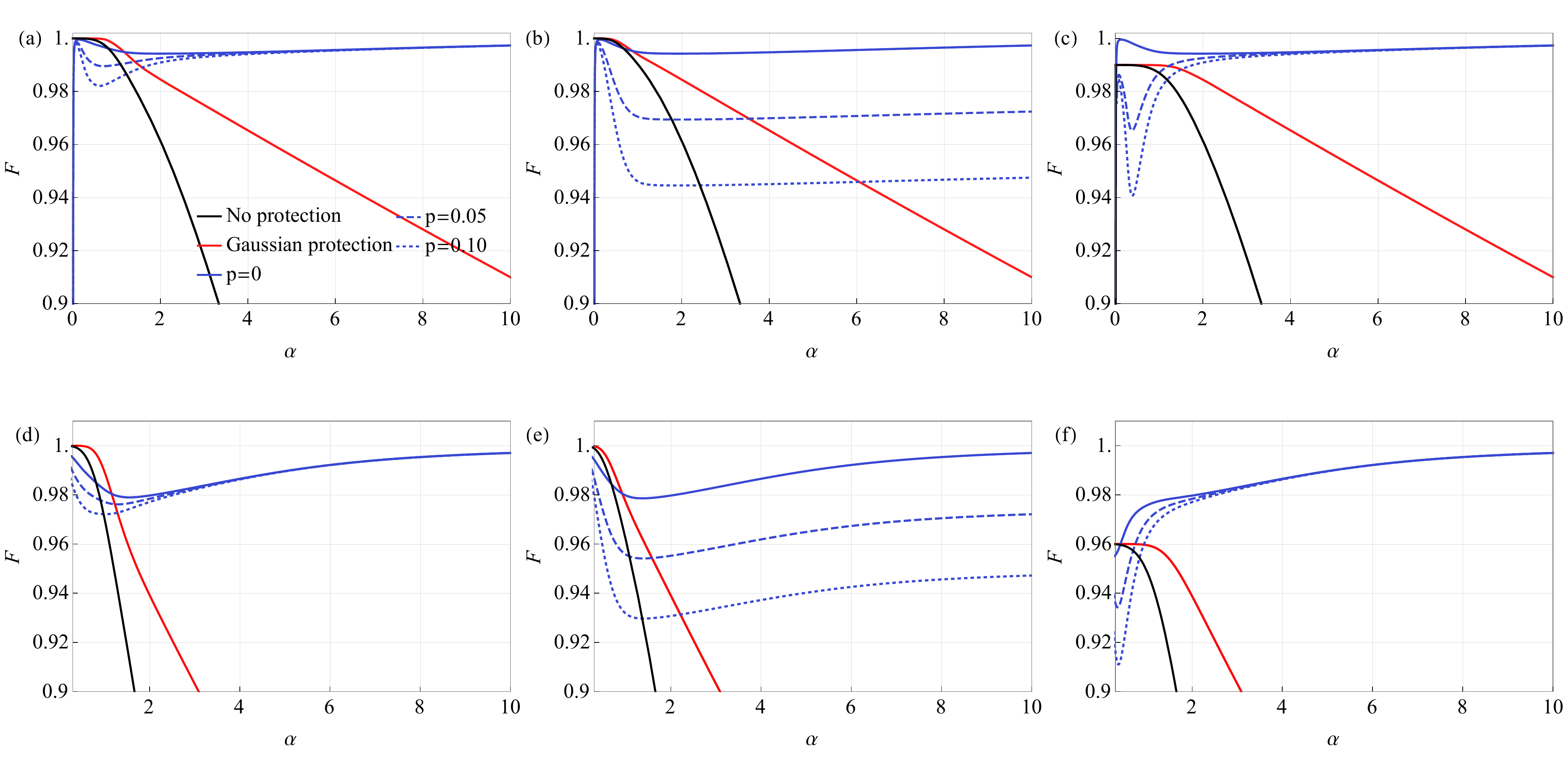}
\caption{
Fidelity vs. $\alpha$ under the different levels of loss for various 2-SCSs. 
Our scheme protects all states better than the optimal Gaussian protection for a large $\alpha$.  
We note that the effect of dephasing is negligible for the real coefficients of the  superpositions. 
The loss (a,b,c) for $1-\eta=0.01$ and (d,e,f) for $1-\eta=0.04$, on input states (a,d)  $\ket{\alpha}+\ket{-\alpha}$, (b,e)  $\ket{\alpha}+i\ket{-\alpha}$, and (e,f)   $\ket{\alpha}-\ket{-\alpha}$.
} \label{fig:indfid}
\end{figure}

In Fig.~\ref{fig:indfid}, it is shown how individual 2-SCS states evolve by the noise channels in terms of fidelity. 
Our bypass protocol is surpassing the Gaussian protocol in terms of the fidelity.
Here, we can see that in some cases the dephasing errors are not affecting the fidelities in the limit of large $\alpha$.
These cases correspond to the case where the qubit enters the channel in the eigenstates of $\hat{\sigma}_z$.

\section{Performance on 4-SCSs}
\label{sec:4scs}

\begin{figure}[th]

\includegraphics[width=450px]{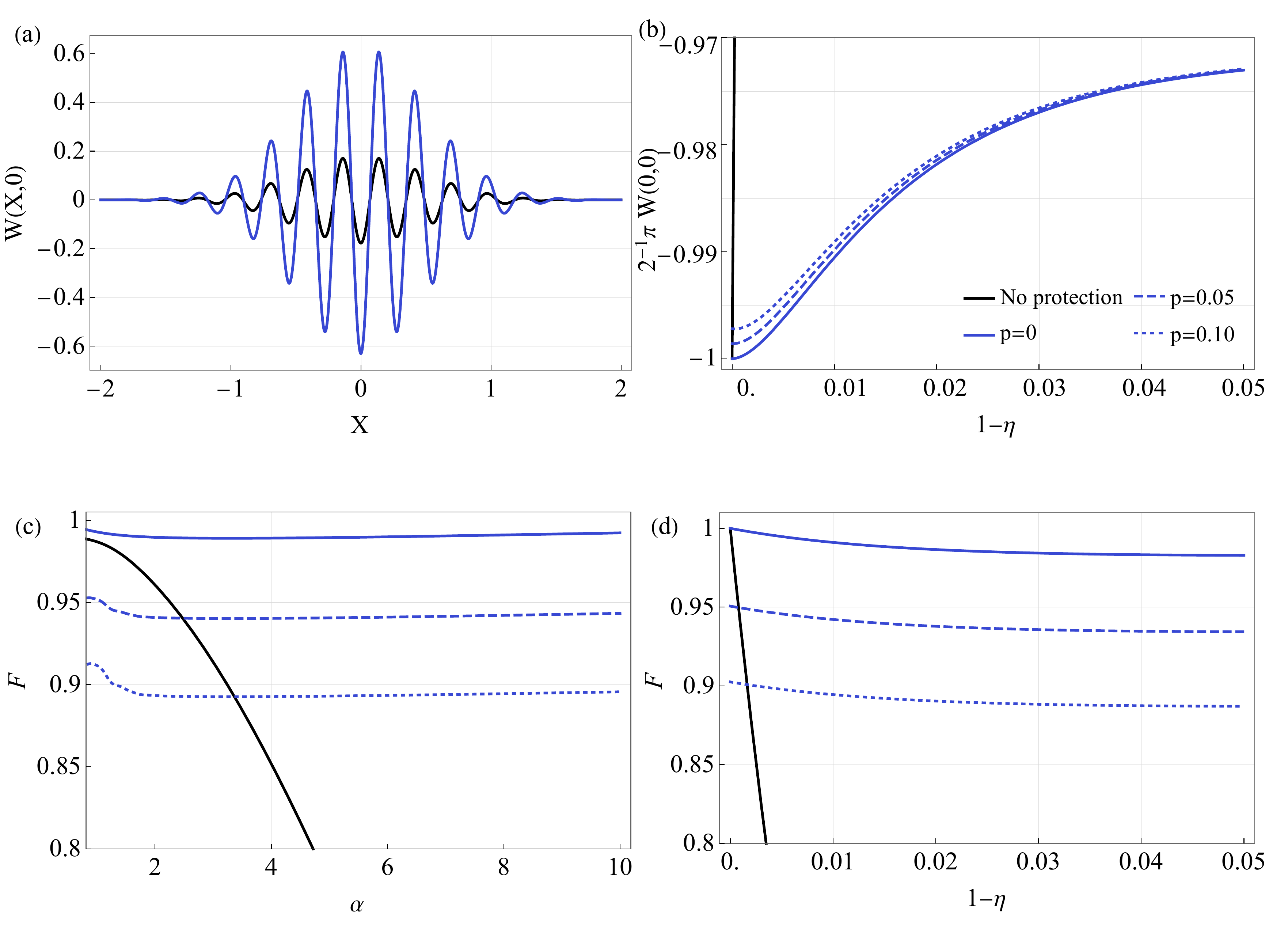}
\caption{  Various measures for the 4-SCS $\ket{\alpha e^{i\pi/4}}-\ket{\alpha e^{3i\pi/4}}-\ket{\alpha e^{5i\pi/4}}+\ket{\alpha e^{7i\pi/4}}$ for $\alpha=8$. 
(a) The cross section of Wigner function along the X-axis for $1-\eta=0.01$,  and  identical curves is obtained along the other axis due to the symmetry.  
The curves for different dephasing factor $p$ are nearly overlapped.
(b) The scaling of the negative peak at the phase space origin vs. $1-\eta$.
 (c) The fidelity vs. $\alpha$ at $1-\eta=0.01$.
(d) The fidelity vs. $1-\eta$ for $\alpha=8$.
} \label{fig:4hcL}
\end{figure}

In Fig.~\ref{fig:4hcL}, we calculated the Wigner functions and channel fidelity for the 4 SCSs under various levels of loss and dephasing. 
We analyzed the  cases corresponding to $\mu(\ket{\alpha e^{i\pi/4}}-\ket{\alpha e^{i5\pi/4}})+\nu(\ket{\alpha e^{i3\pi/4}}-\ket{\alpha e^{i7\pi/4}})$ for a special encodings as in~\cite{LeghtasPRL2013,OfekNature2016CatExperiment}, especially for $\mu=-\nu$.
The Wigner functions of 4-SCSs are comprised of Gaussian peaks of coherent components, the interference fringes of superposition of 2 components, and the interference fringes of superposition of 4 components located near the phase space origin.
The negative peaks of 4-components interference near the phase space origin are well preserved even under the effect of dephasing. 
Similarly as for 2-SCSs, we can also protect 4-SCSs  with an increased fidelity in large $\alpha$ limit against various level of losses even under a modest level of qubit dephasing. 

\section{ Conditional enhancement of protocol by detections}
\label{sec:conddet}


We might consider strategies compromising the deterministic nature in order to enhance the robustness of the protocol.
There are two ways to introduce the projective detections in the protocols, i.e. in the oscillator and the ancillary mode.
If we measure the oscillator after the pre-processing, we can partially filter out the erroneous components with a high success probability.
In (\ref{eq:input}) as in Fig.~\ref{fig:concept} (a), if we detect $\ket{0}\bra{0}$ on APD after the pre-processing $\hat{U}_T$, we are left with nearly target state $i\mu \ket{-_i}-i \nu \ket{+_i}$.
If we detect non-zero photon, we are left nearly with an erroneous mixed state of $i\mu \ket{-_i}-i \nu \ket{-_i}$ and $i\mu \ket{-_i}+i \nu \ket{-_i}$.
If we have access further to a parity detection or corresponding qubit detection through dispersive interaction, we get $i\mu \ket{-_i}-i \nu \ket{+_i}$ for even photon number outcomes and $i\mu \ket{-_i}+i \nu \ket{-_i}$ for odd outcomes.
After a conditional operation $\hat{\sigma}_x$ or $\hat{\sigma}_z$  depending on the outcomes, we get the target state from both states.


The projective qubit detection of the input qubit state after the entire protocol  can additionally filter out the error by loss and dephasing, and the protective power is enhanced. 
The channel fidelity can be improved, e.g. from approximately $0.98$ to $1-5\times 10^{-5}$ for $\alpha=2$ at loss $1-\eta=0.04$ in the absense of dephasing. 
The success probability can be close to 1 if the loss is low, e.g. $99.5\%$ in this example.  
This filtering effect can appear for the other cases similaly, e.g. 4 SCS or ECSs.  

\section{ Hybrid error correction of dephasing using Rabi coupling}
\label{sec:AppB}

We can actively correct the dephasing error in the qubits using Rabi coupling with ancillary CV modes as follows.
Suppose the qubit is in a state $c_1 \ket{+}+c_2 \ket{-}$. 
A Rabi gate $\exp[-i\sqrt{2} \beta \hat{\sigma}_x \hat{P}_1]$ with a small $\beta$ with ancillary oscillator mode in vacuum $\ket{0}_1$ creates an entangled form $c_1 \ket{+}\ket{\beta}+c_2 \ket{-}\ket{-\beta}$. 
We can ignore the loss in the oscillator if $\beta$ is small. 
Now, if qubit state suffers a dephasing $\hat{\sigma}_z$, we have a state $c_1 \ket{-}\ket{\beta}+c_2 \ket{+}\ket{-\beta}$ which corresponds to a bit-flip error of qubit in $\ket{\pm}$ basis. 
Now we apply another Rabi gate $\exp[-i\sqrt{2} \beta \hat{\sigma}_x \hat{P}_2]$ on the qubit and  a second ancillary oscillator in vacuum $\ket{0}_2$ after the channel. 
Then the state without error becomes $c_1 \ket{+}\ket{\beta}_1\ket{\beta}_2+c_2 \ket{-}\ket{-\beta}_1\ket{-\beta}_2$, and  the erroneous state becomes $c_1 \ket{-}\ket{\beta}_1\ket{-\beta}_2+c_2 \ket{+}\ket{-\beta}_1\ket{\beta}_2$ with dephasing.
Now a phononic 50:50 beam splitter interaction is applied between the ancillary modes which transforms the states as follows:
\begin{align}
&\hat{U}_{BS}(c_1 \ket{+}\ket{\beta}_1\ket{\beta}_2+c_2 \ket{+}\ket{-\beta}_1\ket{-\beta}_2)=c_1 \ket{+}\ket{\sqrt{2}\beta}_1\ket{0}_2+c_2 \ket{+}\ket{-\sqrt{2}\beta}_1\ket{0}_2. \nonumber\\
&\hat{U}_{BS}(c_1 \ket{-}\ket{\beta}_1\ket{-\beta}_2+c_2 \ket{+}\ket{-\beta}_1\ket{\beta}_2)=c_1 \ket{-}\ket{0}_1\ket{-\sqrt{2}\beta}_2+c_2 \ket{+}\ket{0}_1\ket{\sqrt{2}\beta}_2.
\end{align}
Now we notice that these two states can be  distinguished:  when the dephasing  occured, the phonons are found only in the mode 2, while without dephasing only in the mode 1. 
Therefore, we can  detect in which mode are phonons, and depending on the outcome apply a correction operation on the qubit.
Note that the number of redundant oscillators affects the error rate and the success probability of this protocol.
Other types of error such as bit-flip error $\hat{\sigma}_x$ can be similarly achieved by using the same formalism with a different Rabi coupling  with the substition of $\hat{\sigma}_x \rightarrow\hat{\sigma}_z$.

An equivalent strategy also exist: the intermediate entangled state in $c_1 \ket{+}\ket{\beta}+c_2 \ket{-}\ket{-\beta}$ becomes after another Rabi gate $\exp[-i\sqrt{2} \beta \hat{\sigma}_x \hat{P}]$ as $c_1 \ket{+}\ket{2\beta}+c_2 \ket{-}\ket{-2\beta}$. 
On the contrary, for the erroneous intermediate state in $c_1 \ket{-}\ket{\beta}+c_2 \ket{+}\ket{-\beta}$, after the same Rabi gate,  we have $c_1 \ket{-}\ket{0}+c_2 \ket{+}\ket{0}$ due to the opposite shift. 
If some phonons are detected, we will know that no dephasing occured, and vise versa. 
These protocol requires phononic detectors that are currently challenging, but can be achieved with complex Rabi circuits and qubit detections.


\section{JC interactions and squeezings}
\label{sec:JCsqueezing}

JC interactions are atom-light interaction similar to Rabi gates, but more available in higher frequency experiments. 
A JC interaction alone cannot replace Rabi gates for our purposes as it lacks the counter-rotating term, but there exist methods which can be derived using JC interactions.
A digital simulation of Rabi gate has been realized by combining JC interactions with a synthetic anti-JC interactions by a Suzuki-Trotter formula~\cite{PedernalesSciRep2015}. Alternatively, the JC interaction can be expressed as the combination of two non-commuting Rabi gates as
\begin{align}
&\exp[i\kappa (\hat{\sigma}_+ a+\hat{\sigma}_- a^\dagger)]=\exp[i\kappa \{\frac{\hat{\sigma}_x+i \hat{\sigma}_y}{2} a+\frac{\hat{\sigma}_x-i \hat{\sigma}_y}{2} a^\dagger\}]=\exp[i2^{-1/2}\kappa(\hat{\sigma}_x \hat{X}-\hat{\sigma}_y \hat{P}) ].\label{JCdecomp}
\end{align} 
Therefore, a Rabi gate can be achieved by removing one of the Rabi terms in JC interaction.

In-line squeezing helps our scheme in many ways. 
First, squeezing can be combined with our scheme to reduce the effect of  loss by reducing the number of excitations in the oscillator as in ~\cite{FilipPRA2013GaussianAdaptation,
LeJeannicPRL2018}. 
Second,  squeezing increases the strength of Rabi gates from $T$ to $T e^{2r}$ as $S[r]\exp[i T \hat{\sigma}_x \hat{X}]S[-r]=\exp[i T e^{2r}\hat{\sigma}_x \hat{X}]$, so the accessible strength of Rabi gate can be increased. 
Third, the digital simulation of  Rabi gates by JC interactions can be improved by a squeezing without accessing rapid qubit rotations:
sandwiching a JC interaction with squeezing and anti-squeezing, we suppress one of the  Rabi terms, and obtain
$S[r]\exp[i2^{-1/2}\kappa(\hat{\sigma}_x \hat{X}-\hat{\sigma}_y \hat{P}) ]S[-r]=\exp[i2^{-1/2}\kappa(\hat{\sigma}_x e^r \hat{X}-\hat{\sigma}_y e^{-r} \hat{P}) ]$.
This single JC operation together with squeezing and anti-squeezing therefore can play the composite Rabi gate $\hat{U}_T$ alone.
Furthermore, by the rotations in the qubit, we get a counter-rotating term $\exp[i \pi \hat{\sigma}_x/2]\exp[i2^{-1/2}\kappa(\hat{\sigma}_x e^r \hat{X}-\hat{\sigma}_y e^{-r} \hat{P}) ]\exp[-i \pi \hat{\sigma}_x/2]=\exp[i2^{-1/2}\kappa(\hat{\sigma}_x e^r \hat{X}+\hat{\sigma}_y e^{-r} \hat{P}) ]$.
When $r$ is large, these two terms can be combined to achieve a single Rabi gate $\exp[i2^{1/2}\kappa\hat{\sigma}_x e^r \hat{X}]$ efficiently.


For the oscillators where in-line squeezing is unavailable, squeezing can be synthesized by alternating  non-commuting Rabi gates $\exp[i \tau \hat{\sigma}_x \hat{X}]$ and $\exp[i \tau \hat{\sigma}_y \hat{P}]$  as 
\begin{align}
\exp[i \tau \hat{\sigma}_x \hat{X}]\exp[i \tau \hat{\sigma}_y \hat{P}]\exp[-i \tau \hat{\sigma}_x \hat{X}]\exp[-i \tau \hat{\sigma}_y \hat{P}]\approx \exp[-2i\tau^2\hat{\sigma}_z (\hat{X}\hat{P}+\hat{P}\hat{X})/2]\exp[i\tau^2\hat{\sigma}_z].\label{eq:engsqz}
\end{align}
Now when the qubit ancilla is in a ground state, this operation is reduced to a single mode squeezing operation.
These synthetic squeezings can be used to enhance the strength of JC interactions in return.
Alternatively, we can use the squeezed state~\cite{Hastrup2020squeezing} for the ancilla to induce squeezing on the target mode.

\section{ Enhancement by multiple Rabi gates}
\begin{figure}[th]
\includegraphics[width=400px]{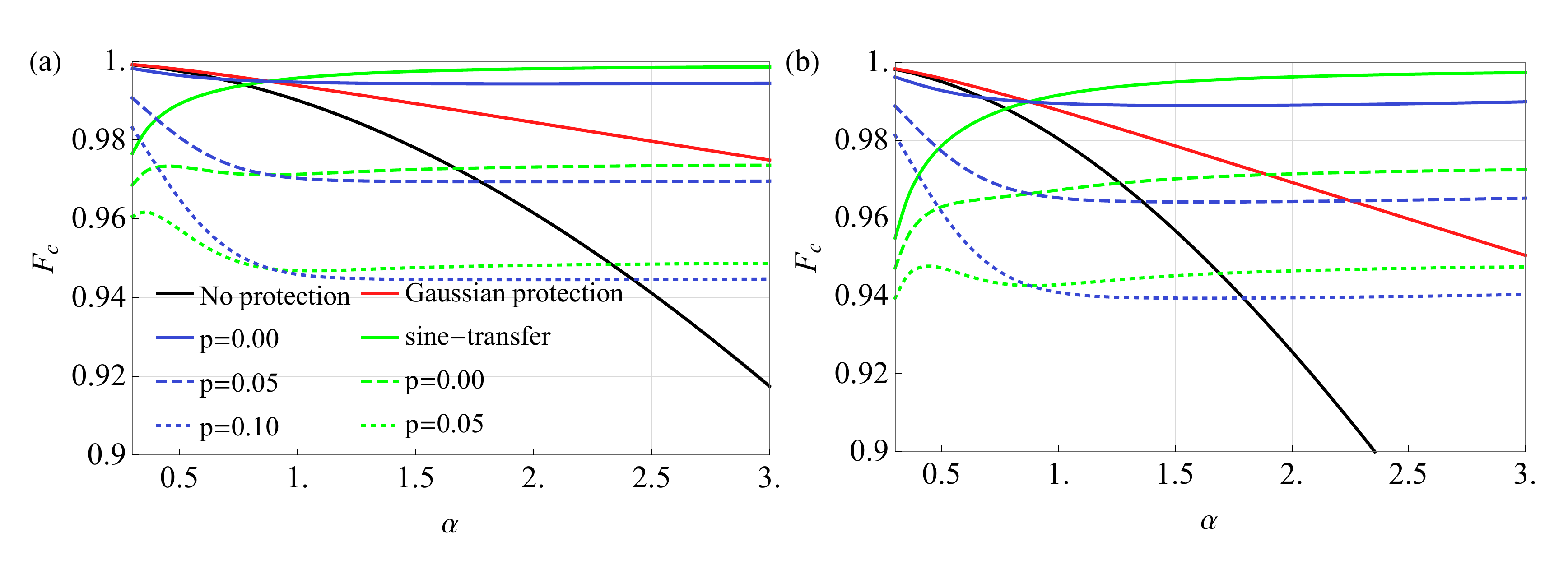}
\caption{ Channel fidelity vs. $\alpha$ of general 2-SCSs under losses (a) $1-\eta=0.01$ and (b) $0.02$, and dephasings $p=0.00,0.05,0.10$.
Using multiple gates, the fidelity is improved, especially under a heavy loss.
 } \label{fig:sine}
\end{figure}

The overall performance of the protocol can be improved further if a constraint in the number of available Rabi gates is relaxed,  by engineering a nonlinear Rabi gate $\hat{U}_D^{V}=\exp[i\hat{\sigma}_x V(\hat{X})]$ with a nonlinear potential $V(\hat{X})$.
An ideal controlled qubit rotation is written as $\hat{U}_D^{\mathrm{sign}}=\exp[i \pi/4 \hat{\sigma}_x \mathrm{sgn}[\hat{X}]]$, where $\mathrm{sgn}[\hat{X}]$ is the sign function. 
 By  this interaction, the qubit ancilla in $\ket{g}$ is rotated perfectly to  $\ket{-}_i$ for the oscillator states lying in right half-plane of phase space, and to $\ket{+}_i$ for the oscillator states in the left half-plane. 
 A variant of Rabi gate  $\hat{U}_D^{\mathrm{tanh}}=\exp[i \pi/4 \hat{\sigma}_x \tanh[\hat{X}]]$ acts nearly as  $\hat{U}_D^{\mathrm{sign}}$ for the states with a large excitation. 
 For large values of quadrature variable $X$, the functional $\tanh[\hat{X}]$ takes a value of $1$ regardless of the value of $X$.
However, $\hat{U}_D^{\mathrm{tanh}}$ is a highly nonlinear operation which is difficult to be implemented. 
A polynomial  Rabi gate $U'_1=\exp[i \epsilon' \hat{\sigma}_x\hat{X}^{k}]$ with $\epsilon' (\sqrt{2}\alpha)^j=\pi/4$ and odd $k$ generates qubit-oscillator entanglement in a similar way as for $k=1$, but the state in the oscillator is heavily disturbed due to the high nonlinearity of the interaction degrading the performance.
 In contrast, another advanced nonlinear Rabi gate has an optimal feasibility and performance:  $\hat{U}_D^{\mathrm{sin}}=\exp[-i \frac{\pi}{4} \hat{\sigma}_x \sin[\frac{\pi}{2\sqrt{2}\alpha} \hat{X}]]$.
This sine-Rabi gate can be approximately achieved by a sequence of three Rabi gates and qubit rotations as in ~\cite{ParkNJP2018}:
\begin{align}
\exp[-i  t_1\hat{X} \hat{\sigma}_y] \exp[i  t_2\hat{\sigma}_z]\exp[2i t_1\hat{X} \hat{\sigma}_y]\exp[-i  t_2 \hat{\sigma}_z]\exp[-i  t_1\hat{X} \hat{\sigma}_y]\approx\exp[i 2 t_2\hat{\sigma}_x\sin[2 t_1\hat{X}]]
\end{align}
at $t_1=\frac{\pi}{4\sqrt{2}\alpha}$ and $t_2=-\pi/8$, where the approximation is good for small $t_{1,2}$. 
This gate acts  very precisely like a controlled qubit rotation $\hat{U}_D^{\mathrm{sign}}$ on the state whose peaks in wavefunction are centered around $X=\pm\sqrt{2}\alpha$ as in Fig.~\ref{fig:sine}, showing a larger enhancement in the protocol especially when the loss is heavy.
This result implies further that individual optimal protective schemes might exist for different types of oscillator state.

\section{ protocols for non-conventional superposition of coherent states }
\label{sec:generalcat}
We now briefly describe how other superposition of coherent states can be protected by a bypass strategy.
We start from an illustrative example of 3-SCS states $\ket{\alpha}+\ket{\alpha e^{i 2\pi/3}}+\ket{\alpha e^{i 4\pi/3}}$, where the three coherent peaks are forming a triangle in phase space.
These peaks can be shifted to be located at the vertices of a rectangle.
First, we displace this state by $\hat{D}[-\alpha/4]$ to make the peaks located symmetrically about the  P-axis of phase space. 
After a Rabi gate $\exp[i \epsilon \hat{\sigma}_x \hat{X}]$ with $\epsilon=\pi/3\sqrt{2}\alpha$, an entangled state is formed approximately as $c_1\ket{+_i}\ket{\alpha-\alpha/4}+c_2\ket{-_i}\ket{\alpha e^{i 2\pi/3}-\alpha/4}+c_3\ket{-_i}\ket{\alpha e^{i 4\pi/3}-\alpha/4}$ with some coefficients $c_j$.
Applying another Rabi gate $\exp[-i \frac{\sqrt{6}}{4}\hat{\sigma}_y\hat{X}]$, the total state becomes $c_1\ket{+_i}\ket{3\alpha/4-i\sqrt{3}\alpha/4}+\ket{-_i}c_2\ket{-3\alpha/4-i\sqrt{3}\alpha/4}+\ket{-_i}c_3\ket{-3\alpha/4+i3\sqrt{3}\alpha/4}$ with coherent peaks located  symmetricaly along the X-axis.
Now applying $\hat{D}[-\sqrt{3}\alpha/4]$, this state becomes  $c_1\ket{+_i}\ket{3\alpha/4-i\sqrt{3}\alpha/2}+\ket{-_i}c_2\ket{-3\alpha/4-i\sqrt{3}\alpha/2}+\ket{-_i}c_3\ket{-3\alpha/4+i\sqrt{3}\alpha/2}$ where the peaks are now located at the vertices of a rectangle, and thus the protective scheme of 4-SCS can finally be applied with use of an additional qubit ancilla.

Another possible extension is the superposition of coherent states located on a line~\cite{ShuklaPRA2019}, such as $\ket{-3\alpha}+\ket{-\alpha}+\ket{\alpha}+\ket{3\alpha}$.
Applying $\exp[i \frac{\pi}{4\sqrt{2}\alpha}\hat{\sigma}_y \hat{X}]$ with the qubit in the ground state, an entangled state generated is $c_{-3\alpha}\ket{-}\ket{-3\alpha}+c_{-\alpha}\ket{+}\ket{-\alpha}+c_{\alpha}\ket{-}\ket{\alpha}+c_{3\alpha}\ket{+}\ket{3\alpha}$.
The information encoded into qubit can again be used to bring the coherent peaks closer to the phase space origin.
Applying $\exp[i \sqrt{2}\alpha \hat{\sigma}_x \hat{P}]$, the state will evolve into $c'_{-3\alpha}\ket{-}\ket{-2\alpha}+c'_{-\alpha}\ket{+}\ket{-2\alpha}+c'_{\alpha}\ket{-}\ket{2\alpha}+c'_{3\alpha}\ket{+}\ket{2\alpha}$.
We can now use the formalism of 2-SCS bypass protocol to transform the oscillator into vacuum.

A non-trivial question  is if we can improve the protocol using more resources.
We note that a 2-SCS $\mu\ket{\alpha}+\nu\ket{-\alpha}$ can be understood as a special case of superposition of 4-SCS with zero weight in the 2 other vertices of a rectangle $\ket{-\alpha}$ and $\alpha$. 
This connection can be made clear by a Gaussian transformation $R[-\pi/4]\hat{D}[i\alpha]$ to $\mu\ket{\sqrt{2}\alpha}+\nu\ket{\sqrt{2}i\alpha}$, which is a special case of $\mu_1\ket{\sqrt{2}\alpha}+\mu_2\ket{\sqrt{2}i\alpha}+\mu_3\ket{-\sqrt{2}\alpha}+\mu_4\ket{-\sqrt{2}i\alpha}$. 
Here we note that the absolute amplitude of the states has been increased from $\alpha$ to $\sqrt{2}\alpha$. 
Therefore, using the 4-SCS bypass formalism with 2 ancillas, we can expect an enhanced protection.

The trick of using more ancillas for enhanced protection  can be applied to other n-SCSs using proper controlled rotations and displacements as follows.
For a general n-SCS, we can think of the following example.
Let us assume the n-SCS as the coherent states aligned at the grid-like locations.
As this grids can be viewed as the alignment in 2-dimension, each axis can be treated separately as in the case of superposition of coherent states on a line.
This idea can also work to an arbitrary input unknown states~\cite{JacobDigitalizer}. 

\bibliography{testbibtex}{}


\end{document}